\documentclass[10pt,a4paper]{article}
\usepackage{amsmath,amssymb}
\usepackage{graphicx}
\usepackage{subfigure}
\usepackage{fancyhdr}
\usepackage[textwidth=14cm,textheight=22cm]{geometry}
\begin{document}
\title{\large{Local and Nonlocal Thermal Transport Simulation}}
\author{\large{\textbf{Kaifeng Chen}}}
\date{}
\maketitle
\clearpage
\tableofcontents
\clearpage
\section{\Large{\textbf{Introduction}}}
In the field of Inertial Confinement fusion(ICF),laser-produced plasmas generally fall between collisionless and collisional regimes and could not simply be considered as being in local thermal equilibrium because of small size and short duration of the plasmas.\\[0.1cm]
When a plasma is weakly coupled, i.e. the Coulomb logarithm $ln\Lambda$ of the plasma is much larger than 1, transport process is mainly governed by the multiple small-angle Coulomb scattering between charged particles. and can be accurately described by Fokker-Planck(FP) equation.By comparison, a classical model named Spitzer-Harm theory has been proposed for long time. However,steep electron temperature gradient around critical density make the classical Spitzer-Harm theory failure to describe the electron thermal conduction in this region. Since energetic electrons carrying the heat flux have a long mean-free path that in the same order of or even longer than the scale length of the electron temperature, the electron heat flux is determined by the electron temperature distribution rather than the local temperature gradient. The electron transport becomes non-local.\\[0.1cm]
Much of the progress in the Fokker-Planck simulation of a plasma has been possible following the pioneer work of CHANG and COOPER. Based upon previous work, EPPERLEIN has advanced an implicit and conservative difference scheme for the Fokker-Planck equation. In this work, we apply Epperlein's scheme to develop a Fokker-Planck code written in C language by Doc. Bin Zhao.
\section{\Large{\textbf{Fokker-Planck equations and numerical scheme}}}
The electron Fokker-Planck equation for a high-Z laser produced plasma can be written as:
\begin{equation}
 \frac{\partial f}{\partial t}+v\cdot \frac{\partial f}{\partial r}-\frac{e}{m_e}E\cdot\frac{\partial f}{\partial v}=C_{ee}(f)+C_{ei}(f)
\end{equation}
$C_{ee}$ is the electron-electron collision operator, and $C_{ei}$ is the Lorentz electron-ion collision operator. Then distribution function $f(x,v,t)$ can be expanded in a series of Legendre polynomials:
\begin{equation}
 f(x,v,t)=\sum_{l=0}^{\infty}f_l(x,v,t)P_l(\mu)
\end{equation}
where $\mu=v_x/v=cos\theta$ is the direction cosine. In the diffusive approximation, the series (2) is truncated at $N=1$, namely, only $f_0$ and $f_1$ are kept in the expansion, and the time derivative of $f_1$ is neglected. It is easy to obtain the equation for the isotropic part of the distribution function which is coupled with the anisotropic part $f_1$:
\begin{equation}
 \frac{\partial f_0}{\partial t}+\frac{v}{3}\frac{\partial f_1}{\partial x}=\frac{1}{v^2}\frac{\partial}{\partial v}\big(\frac{v^2}{3}\frac{eE}{m_e}f_1\big)+C_{ee}(f_0,f_0)
\end{equation}
For high-Z plasma, the anisotropic part of the electron-electron collision term is negligible in comparison with that of electron-ion collision term. The equation for $f_1$ can be simplified as:
\begin{equation}
 f_1=-\frac{1}{\nu_{ei}}\big(v\frac{\partial f_0}{\partial x}-\frac{eE}{m_e}\frac{\partial f_0}{\partial v}\big)
\end{equation}
where $\nu_{ei}=4\pi N_eZe^4/m_e^2ln\Lambda /v^3$, $Z$ is the ion charge state, $N_e$ is the electron number density. Then we obtain a closed equation of $f_0$.
\begin{equation}
\begin{split}
 \frac{\partial f_0}{\partial v}&=\frac{\partial}{\partial v}\big[\chi(\frac{\partial f_0}{\partial x}-\frac{a}{v}\frac{\partial f_0}{\partial v})\big]+\frac{1}{v^2}\frac{\partial}{\partial v}\\
 &\times\big[\chi(a^2\frac{\partial f_0}{\partial v})-av\frac{\partial f_0}{\partial x}\big]+C_{ee}(f_0,f_0)+S_{IB}
\end{split}
\end{equation}
where $a=eE/m_e.\chi=v^2/3\nu_{ei}$. The electron-electron collision operator $C_{ee}(f_0,f_0)$ is given by
\begin{equation}
 C_{ee}(f_0,f_0)=\Gamma_{ee}\frac{1}{v^2}\frac{\partial}{\partial v}\big[C(f_0)f_0+D(f_0)\frac{\partial f_0}{\partial v}\big]
\end{equation}
where $\Gamma_{ee}=4\pi e^4/m_e^2ln\Lambda =v_T^3/N_e\tau$,$\tau$ is the electron-electron collision time.$v_{\tau}$ is the electron thermal velocity, and $C(f_0)$ and $D(f_0)$ are defined as
\begin{align}
 C(f_0)&=\int_0^vduu^2f_0(u,t)\\
 D(f_0)&=\frac{1}{3v}\int_0^vduu^4f_0(u,t)+\frac{v^2}{3}\int_v^{+\infty}duuf_0(u,t)
\end{align}
The self-consistent eletrostatic field $E$ or equivalently $a$ can be obtained from the quasi-neutrality condition. Taking the zeroth order velocity moment, we obtain the continuity equation for the electron density $N_e$.
\begin{equation*}
 \frac{\partial N_e}{\partial t}+\frac{\partial}{\partial v}\Gamma=0
\end{equation*}
where $\Gamma$ is the electron flux.
\begin{equation}
 \Gamma=-4\pi\int_0^{\infty}v^2dv\chi\big(\frac{\partial f_0}{\partial x}-\frac{a}{v}\frac{\partial f_0}{\partial v}\big)
\end{equation}
The condition of quasi-neutrality requires a divergence-free partial flux $\Gamma$. With the assumption of the immobile ion background, the self-consistent eletrostatic field $a$ can be obtained with the condition $\Gamma=0$,
\begin{equation}
 a=\frac{\int dvv^2\chi(\partial f_0/\partial x)}{\int dvv^2(\chi/v)(\partial f_0/\partial v)}
\end{equation}
With the condition of quasi-neutrality, the second term on the right hand side of Eq.(5),usually called Ohmic term, can be neglected.\\[0.1cm]
On the right hand side of Eq.(5), we add an additional term $S_{IB}$ to include the IB absorption of the laser energy. According to LANGDON's theory, $S_{IB}$ can be written as
\begin{equation}
 S_{IB}(f_0)=\frac{N_eZ\Gamma_{ee}v_o^2}{6}\frac{1}{v^2}\frac{\partial}{\partial v}\big(\frac{1}{v}\frac{\partial f_0}{\partial v}\big)
\end{equation}
In order to ensure better energy conservation, we calculate $D$ directly from the equation
\begin{equation}
\frac{\partial}{\partial v}(vD)=v^2\int_0^{\infty}duuf
\end{equation}
Phase-space differentiation is carried out with on a fixed Cartesian grid in $(x,v)$.The spatial grid is usually uniform and the grid of the velocity space can be feathered with finer resolution at low velocity grid when the laser-heating effect is taken into consideration in the simulation. In velocity space, number conserving boundary conditions are adopted. In order to keep a zero heat flux in the spatial domain boundary, fixed boundary conditions are implemented in spatial space.\\[0.1cm]
My work mainly concentrates on the comparison between classical thermal transport by using Spitzer-Harm theory and the Fokker-Planck simulation. Some examples concerned with different aspects are shown in following sections.
\section{\Large{\textbf{When the initial density has linear profile}}}
\subsection{\large{\textbf{Case 1: $n_e=10^{14}cm^{-3}$}}}
In this example of simulation, the conditions are as follows:
\begin{itemize}
\item no laser heating.
\item density profile: linear from 4ev to 5ev, 4ev corresponds to the left point and 5ev corresponds to the right point.
\item only considering a 1-D model with the length $L=200cm$, the space is divided into 100 girds.
\item time step is $\delta t=1.0\times 10^{-6}s$,which means after $\delta t$ all the parameters of the system will be recalculated. After every $100\times\delta t$, the code generates an output.
\end{itemize}
The simulation results are shown as follows(Figure 1,$t=20000ns$), the dark line represents the result by using Fokker-Planck equation, while the red one represents the result by using classical SH model.
\begin{figure}[!h]
\centering
\includegraphics[height=8cm,width=14cm]{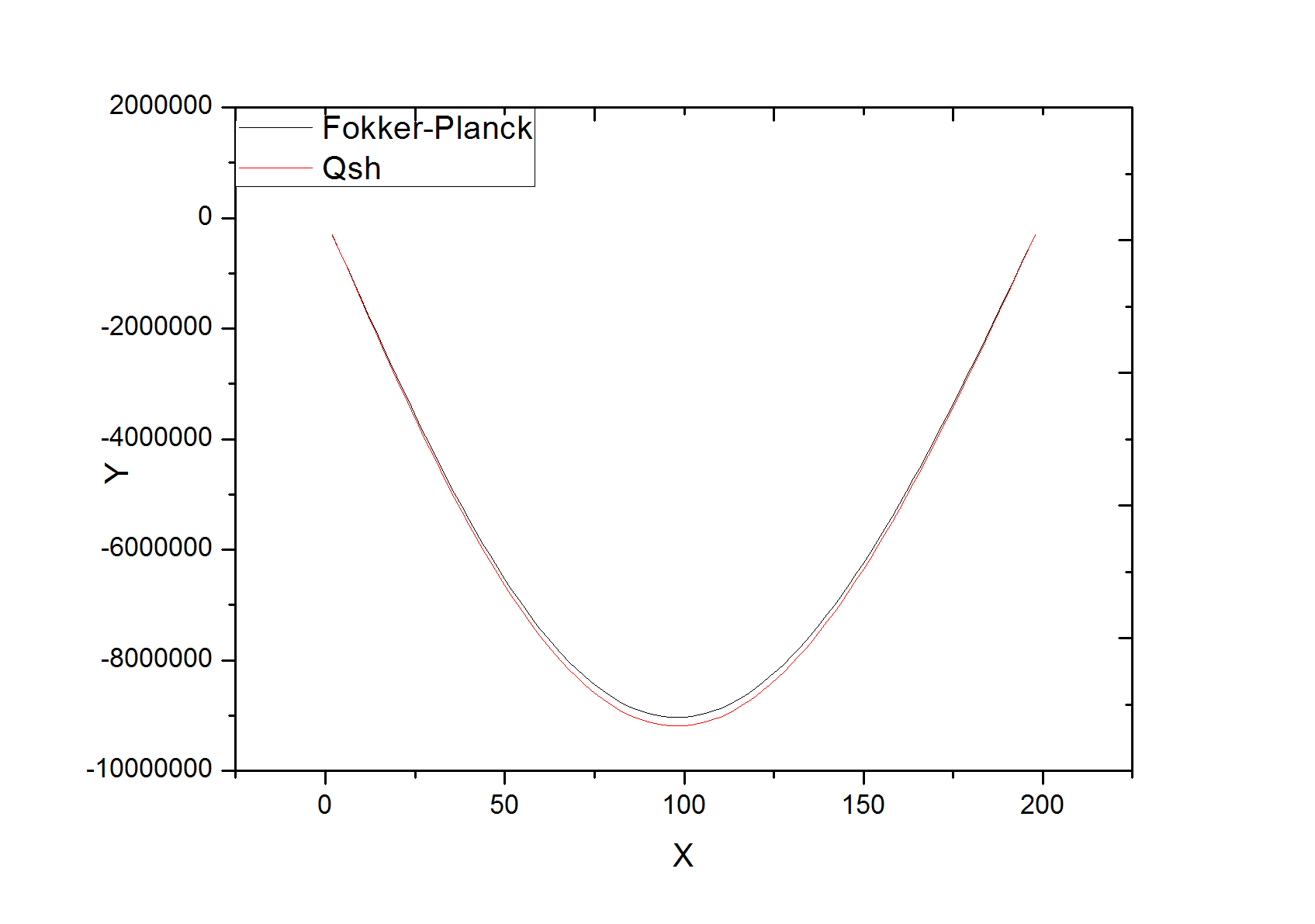}
\caption{Heat Flux}
\end{figure}\\[0.1cm]
The mean free path is $0.256cm$ at this time point,which is much shorter than the width of each grid($2cm$).The electron-electron collision time is $1.22\times 10^{-9}s$.\\[0.1cm]
By examining the results,the two different models bring out nearly the same results. This indicates that under this certain conditions, the classical theory is still applicable.\\[0.1cm]
The following picture[Figure 2] plots the temperature profile at $t=20000ns$.
\begin{figure}[!h]
\centering
\includegraphics[height=8cm,width=14cm]{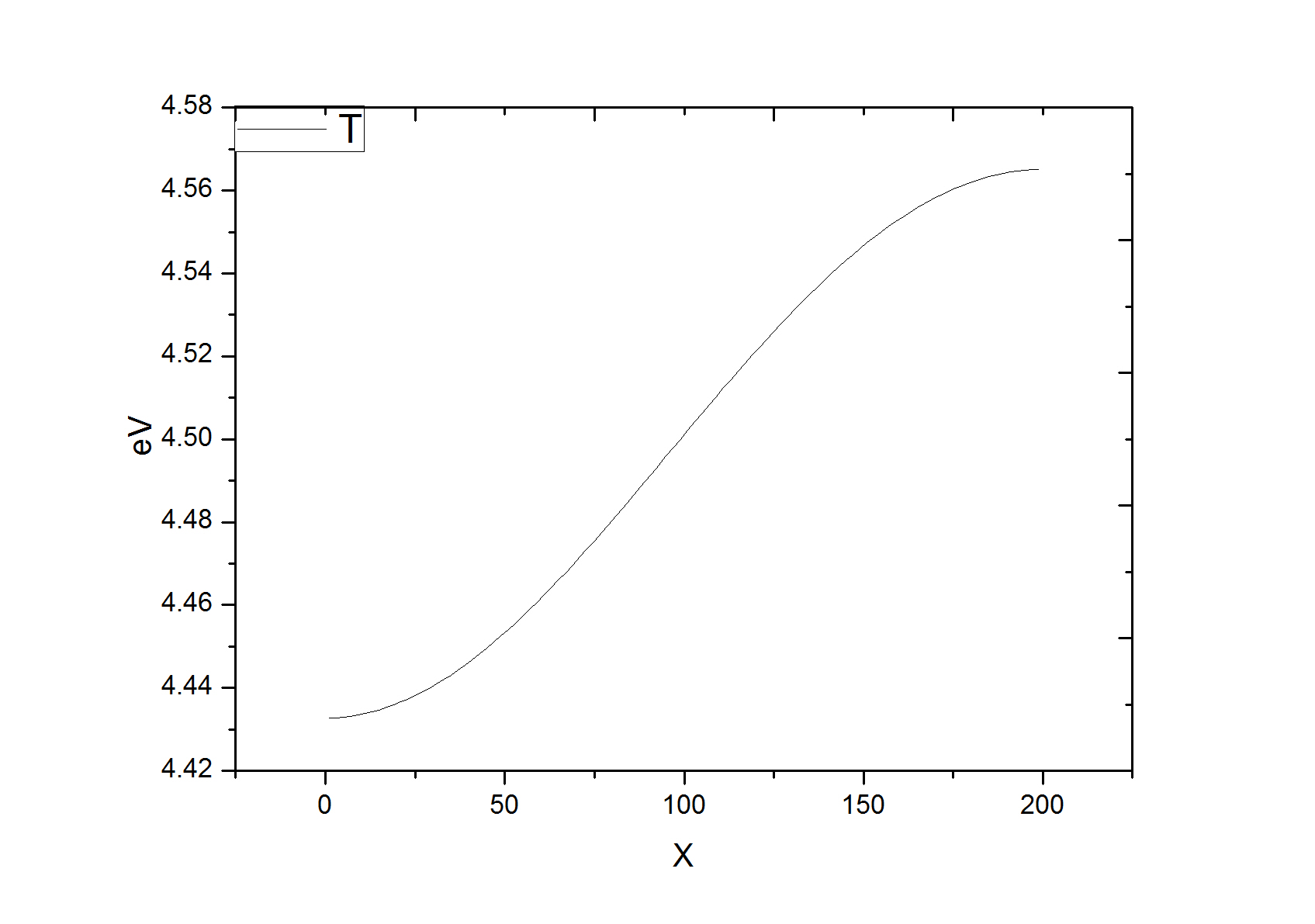}
\caption{Temperature Profile}
\end{figure}
\subsection{\large{\textbf{Case 2: $n_e=10^{11}cm^{-3}$}}}
In this example of the simulation, the mainly difference from case 1 is the electron density. Meanwhile, some simulation parameters have been changed:
\begin{itemize}
\item the density is $10^{11}cm^{-3}$
\item time step is $\delta t=1.0\time 10^{-8}s$
\end{itemize}
The following picture[figure 3] shows the simulation results.From the picture,the two different results show distinct difference in calculating heat flux.This means,in this certain case,due to the low density of electron,the local model should be applied and the classical theory is not valid at all.
\begin{figure}[!h]
\includegraphics[height=8cm,width=14cm]{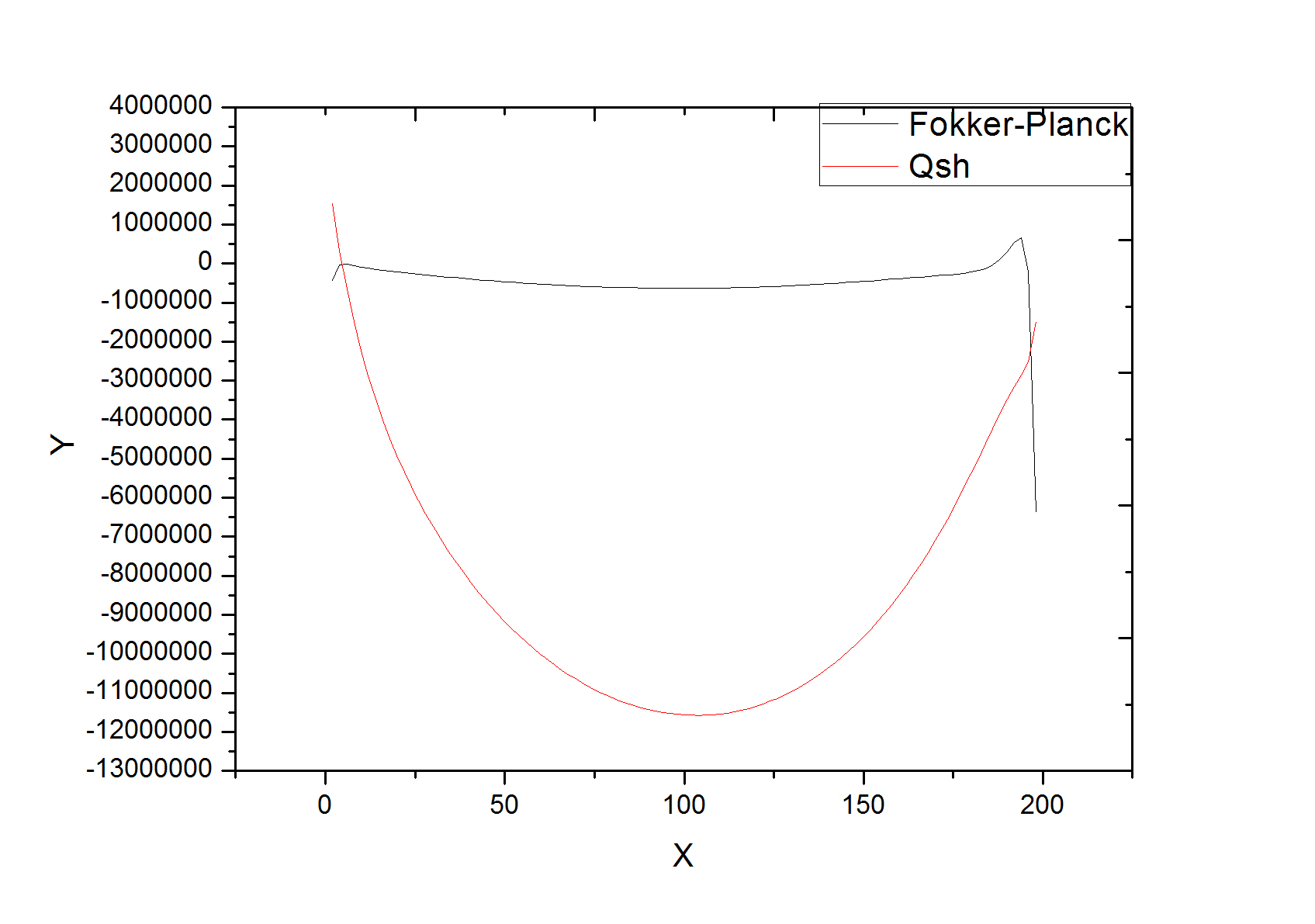}
\caption{Heat Flux}
\end{figure}\\[0.1cm]
The following picture[figure 4] shows the temperature distribution along $x$ axis when $t=2000ns$. It is no doubt that the final temperature will be all uniform in space.
\begin{figure}[!h]
\includegraphics[height=8cm,width=14cm]{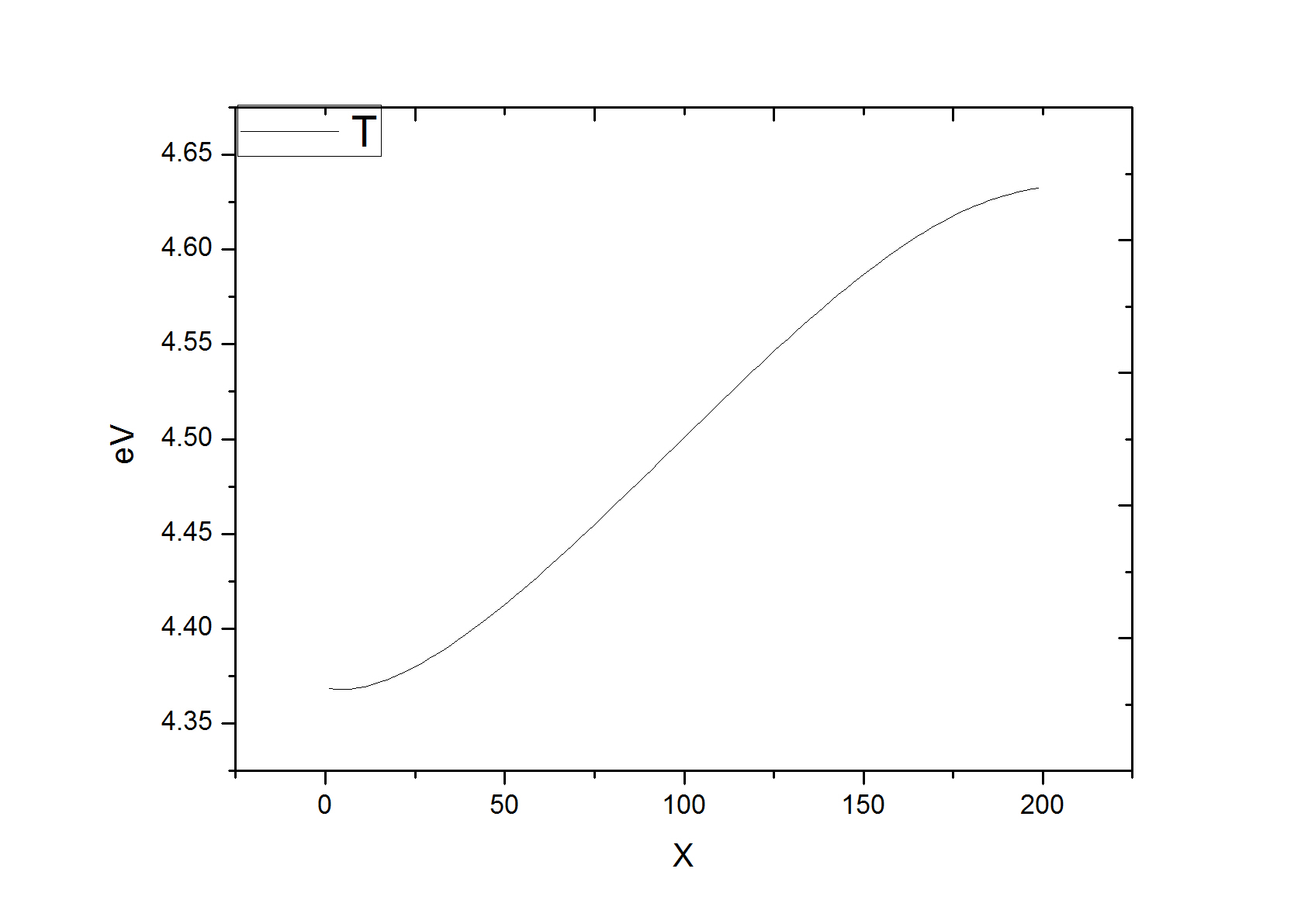}
\caption{Temperature Profile}
\end{figure}
\section{\Large{\textbf{When the classical heat flux remains constant at initial state}}}
When the classical heat flux remains constant, the initial temperature profile will be in a certain mathematical form. The classical thermal transport can be written as:
\begin{equation}
 Q_{sh}=C\cdot T^{5/2}\nabla T
\end{equation}
If heat flux remains a constant,say $C_0$, then we can get:
\begin{equation}
 C\cdot T^{5/2}\nabla T=C_0
\end{equation}
Thus the temperature should be
\begin{equation}
 T=(C_1x+C_2)^{2/7}
\end{equation}
So if given the temperature value of two points along the axis,the $C_1$ and $C_2$ can be solved.\\[0.1cm]
In the simulation,the number of grid is 400,and the number of the grids of velocity space is 80. Here we consider a period as long as 400cm. We set the first 50 girds to be 4ev and the last 40 grids to be 5ev,in order to better analyze the evolution of temperature according to time. Like the above discussions,two examples of $n_e=10^{14}cm^{-3}$ and $n_e=10^{11}cm^{-3}$.
\subsection{\large{\textbf{Case 1: $n_e=10^{14}cm^{-3}$}}}
As for this case,the parameters are listed below:
\begin{itemize}
\item time step is $\delta t=1.0\times 10^{-8}s$.
\item for every 100 steps,the program generates an output.
\end{itemize}
The result presented in Figure 5 when $t=2000ns$.
\begin{figure}[!h]
\centering
\includegraphics[height=8cm,width=14cm]{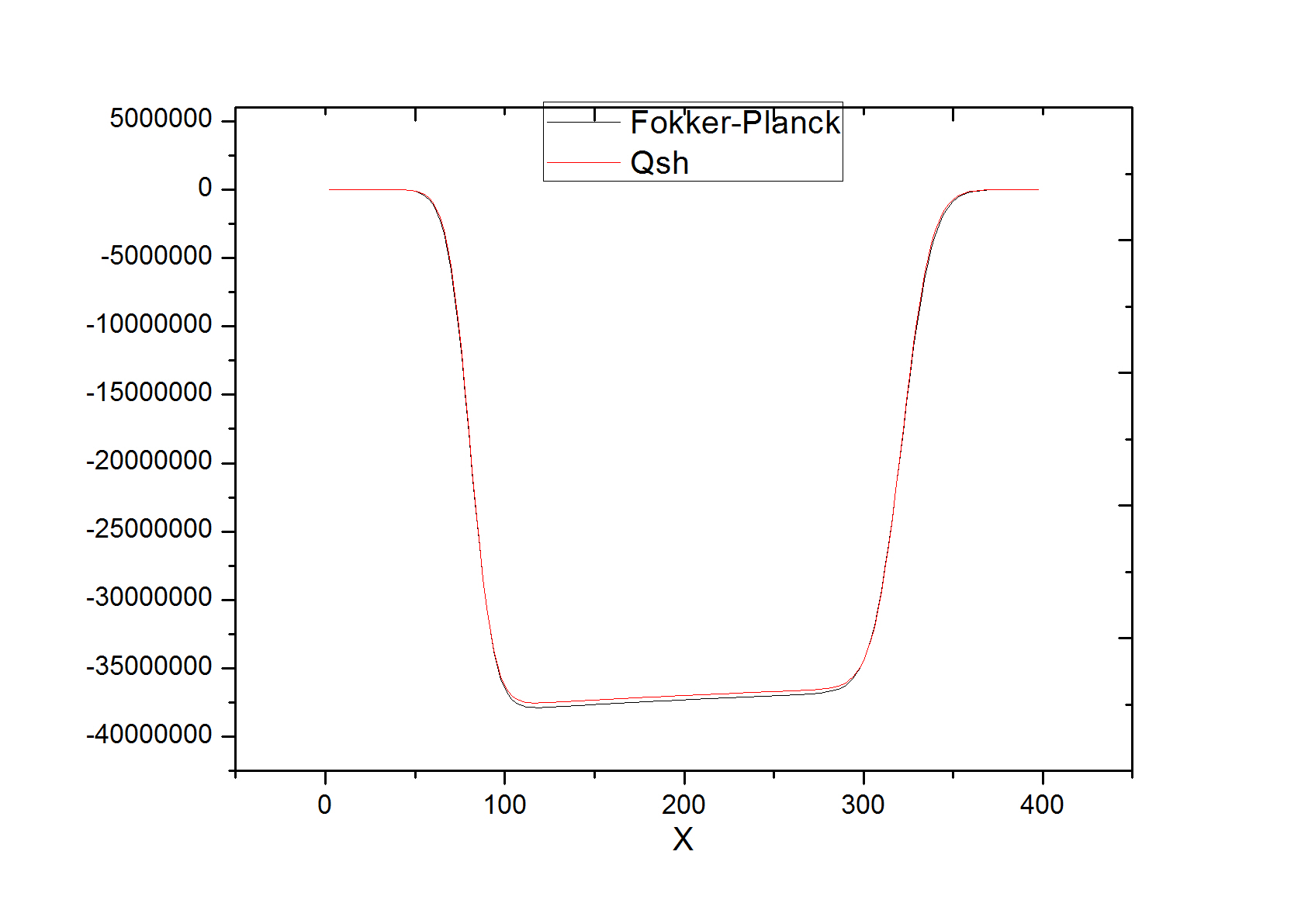}
\caption{Heat Flux}
\end{figure}
The result shows that it is a classical case,in which classical Spitzer-Harm theory can get the right physical result.The calculated mean free path for electrons is $0.256cm$,and the collision time of electrons and electrons is $1.22\times 10^{-9}$.\\[0.1cm]
The following picture gives the temperature profile at this time point.
\begin{figure}[!h]
\centering
\includegraphics[height=8cm,width=14cm]{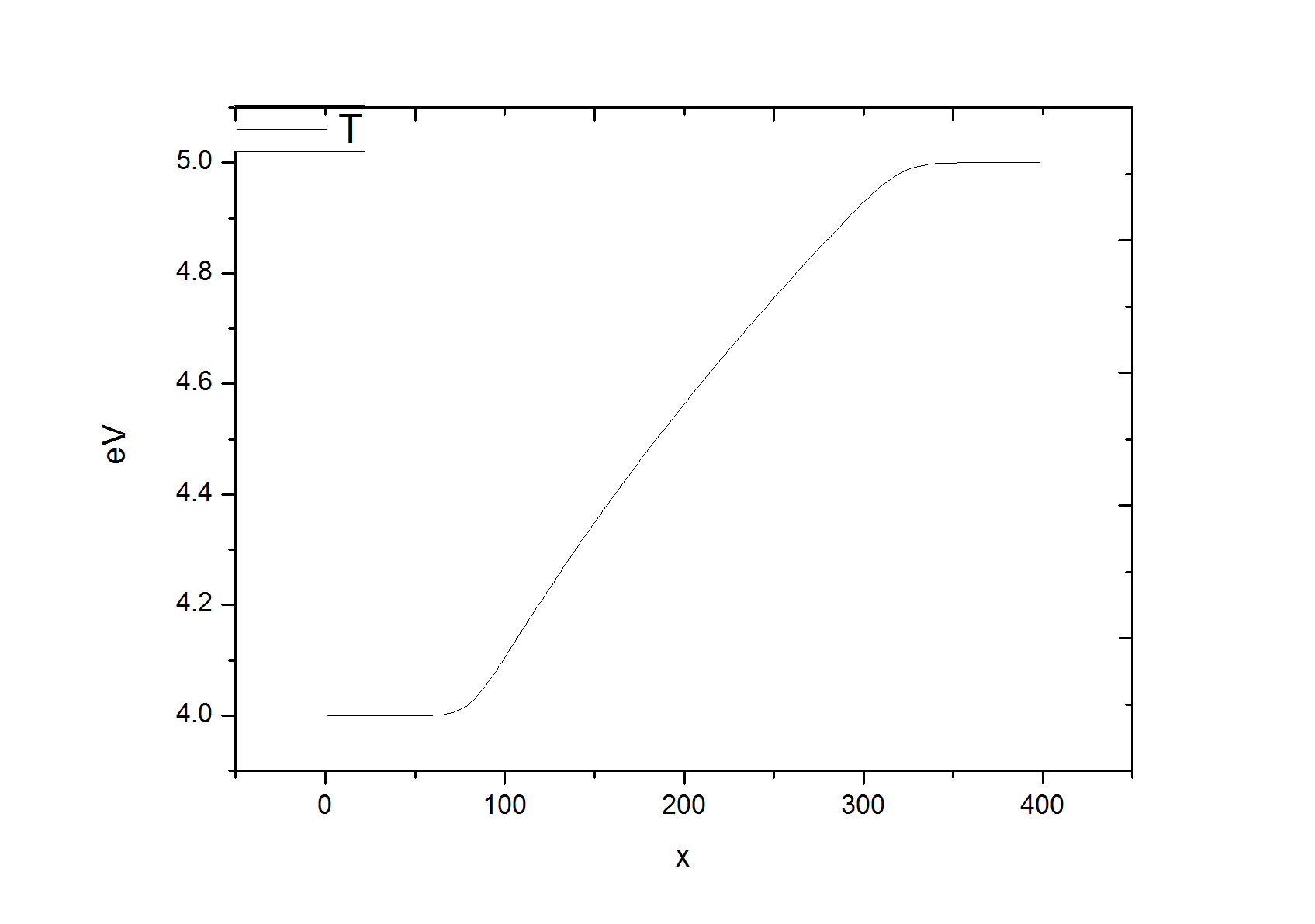}
\caption{Temperature Profile}
\end{figure}
\subsection{\large{\textbf{Case 2: $n_e=10^{11}cm^{-3}$}}}
The mean free path in this case is $181.5cm$. Such long mean free path leads to totally local thermal transport.The time step is set $1.0\times 10^{-10}s$ and for every 200 steps, we get a result.\\[0.1cm]
Figure 7 gives the heat flux when $t=40ns$.
\begin{figure}[!h]
\centering
\includegraphics[height=8cm,width=14cm]{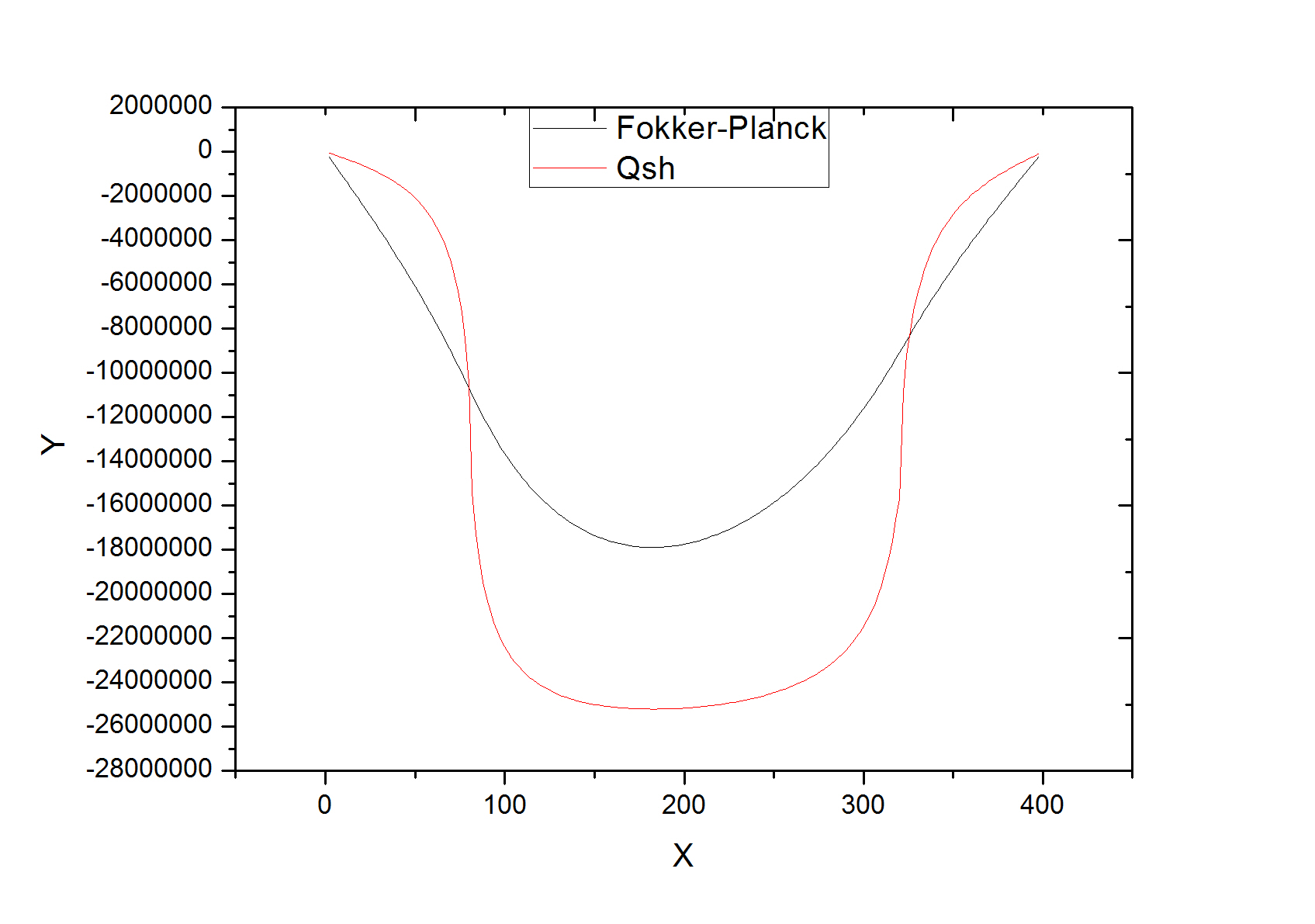}
\caption{Heat Flux}
\end{figure}\\[0.1cm]
As for the temperature profile,it is shown in Figure 8.
\begin{figure}[!h]
\centering
\includegraphics[height=8cm,width=14cm]{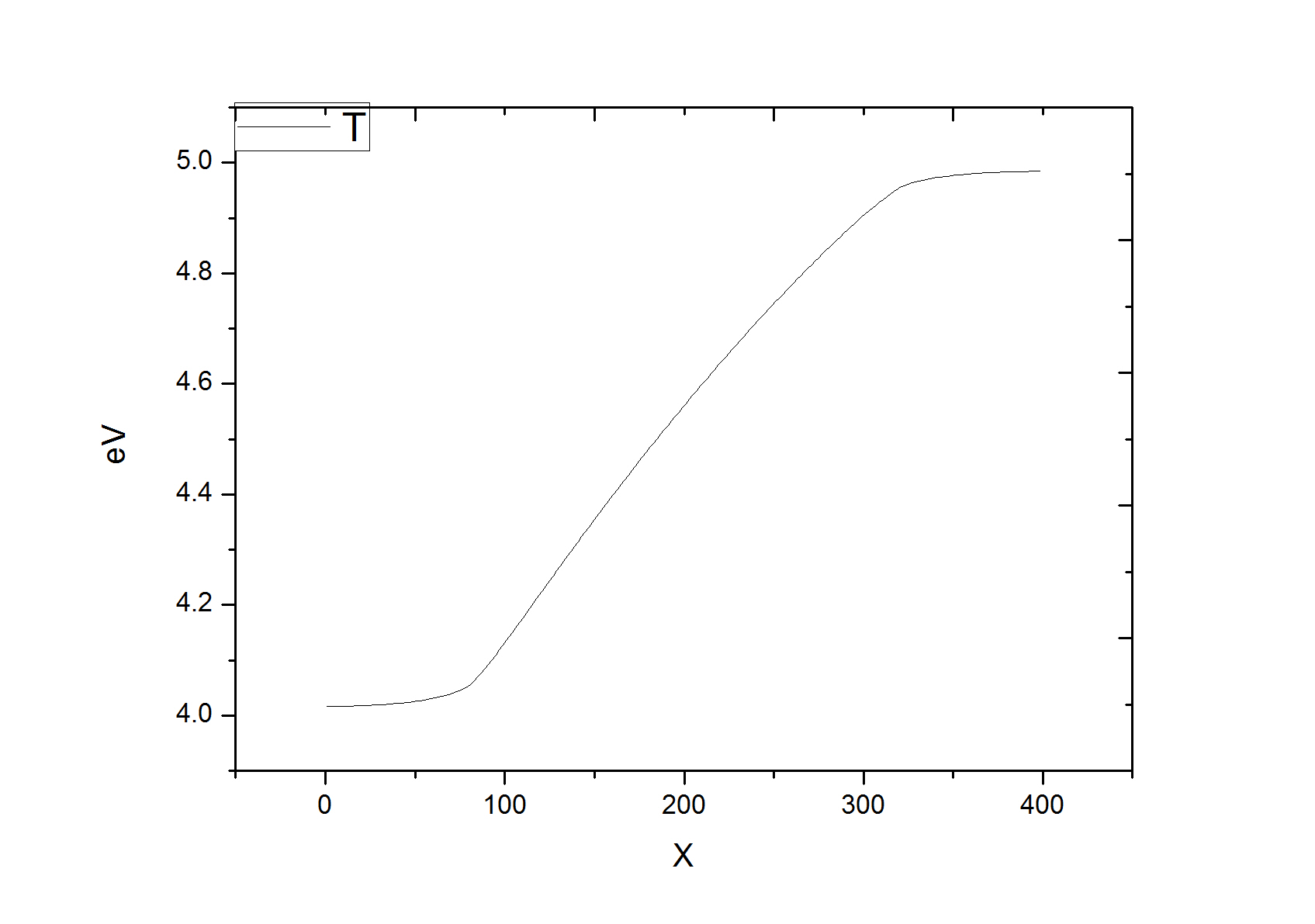}
\caption{Temperature Profile}
\end{figure}
\section{\Large{\textbf{Conclusion}}}
In summary, we develop a Fokker-Planck code to investigate the electron transport process in laser-produced plasma in relevance to inertial confinement fusion. The diffusive approximation is adopted to simplify the Fokker-Planck equation. The equations are numerically implemented by infinite difference method and solved implicity in time.By comparing the classical Spitzer-Harm theory and the Fokker-Planck equation,the physical picture of local transport and nonlocal transport becomes much clearer. When the temperature gradient is deep or the density of electrons is low, the effect of local thermal transport will become dominant which gives total different results. Always is the case that the nonlocal thermal heat flux is less than classical heat flux within the calculated region.However, by now we only talk about a simple example in which there is no laser heating,and this case is common in MCF when the magnetic field is so weak that it can be neglected.As for some complex conditions, further simulations and discussions will be the future work we will do.
\section{\Large{\textbf{References}}}
1.Rosenbluth M N, MacDonald W M, Judd D L.1957,Phys.Rev.,1:107\\[0.1cm]
2.Epperlein E M.1994,Laser Part.Beams.257:12\\[0.1cm]
3.Epperlein E M.1994,J.Comput.Phys.112:291\\[0.1cm]
4.Zhao Bin,Zheng Jian.2008,Plasma Science and Technology,Vol.10,No.1,Feb.
\end{document}